# Block matching algorithm based on Differential Evolution for motion estimation

*Erik Cuevas, *Daniel Zaldívar[1], *Marco Pérez-Cisneros and +Diego Oliva

*Departamento de Electrónica
Universidad de Guadalajara, CUCEI
Av. Revolución 1500, C.P 44430, Guadalajara, Jal, México
{erik.cuevas, [1]daniel.zaldivar, marco.perez}@cucei.udg.mx

+Dpto. Ingeniería del Software e Inteligencia Artificial,
Facultad Informática, Universidad Complutense,
Av. Complutense S/N, 28040, Madrid, Spain
doliva@estumail.ucm.es

**Abstract**

Motion estimation is one of the major problems in developing video coding applications. Among all motion estimation approaches, Block matching (BM) algorithms are the most popular methods due to their effectiveness and simplicity for both software and hardware implementations. A BM approach assumes that the movement of pixels within a defined region of the current frame (Macro-Block, MB) can be modeled as a translation of pixels contained in the previous frame. In this procedure, the motion vector is obtained by minimizing the sum of absolute differences (SAD) produced by the MB of the current frame over a determined search window from the previous frame. The SAD evaluation is computationally expensive and represents the most consuming operation in the BM process. The most straightforward BM method is the full search algorithm (FSA) which finds the most accurate motion vector, calculating exhaustively the SAD values for all elements of the search window. Over this decade, several fast BM algorithms have been proposed to reduce the number of SAD operations by calculating only a fixed subset of search locations at the price of a poor accuracy. In this paper, a new algorithm based on Differential Evolution (DE) is proposed to reduce the number of search locations in the BM process. In order to avoid computing several search locations, the algorithm estimates the SAD values (fitness) for some locations using the SAD values of previously calculated neighboring positions. Since the proposed algorithm does not consider any fixed search pattern or other different assumption, a high probability for finding the true minimum (accurate motion vector) is expected. In comparison to other fast BM algorithms, the proposed method deploys more accurate motion vectors yet delivering competitive time rates.

*Keywords*: Block matching algorithms, motion estimation, differential evolution, fitness approximation.

## 1. Introduction

Virtually all applications of video and visual communication deal with an enormous amount of data. The limited storage capacity and transmission bandwidth available has made digital video coding an important technology. In video coding, the high correlation between successive frames can be exploited to improve coding efficiency, which is usually achieved by using motion estimation (ME). Many ME methods have been studied in an effort to reduce the complexity of video coding, such as block matching (BM) algorithms, parametric-based models [1], optical flow[2] and pel-recursive techniques [3].  Among these methods, BM seems to be the most popular method due to its effectiveness and simplicity for both software and hardware implementations. BM is also widely adopted by various video coding standards, such as MPEG-1 [4], MPEG-2 [5], MPEG-4 [6], H.261 [7] and H.263 [8].

---

[1] Corresponding author, Tel +52 33 1378 5900,  Ext. 7715, E-mail: daniel.zaldivar@cucei.udg.mx

In a BM algorithm, the current frame is divided into non-overlapping macro blocks of $N$x$N$ pixel dimension. For each block, in the current frame, the best matched block within a search window of size $(2W+1)$x$(2W+1)$ in the previous frame is determined, where $W$ is the maximum allowed displacement. The position difference between a template block in the current frame and the best matched block in the previous frame is called the motion vector. A commonly used matching measure is the sum of absolute differences (SAD) which is computationally expensive and represents the most consuming operation in the BM process.

The full search algorithm (FSA) [9] is the simplest block-matching algorithm that can deliver the optimal estimation solution regarding the minimal matching error as it checks all candidates one at a time. However, such exhaustive search and full-matching error calculation at each checking point yields an extremely computational expensive FSA method that seriously constraints real-time video applications.

In order to decrease the computational complexity of the BM process, several BM algorithms have been proposed which are based on the following three techniques: (1) Using a fixed pattern, which means that the search operation is conducted on a fixed subset of the total search window. The Three Step Search (TSS) [10], the New Three Step Search (NTSS) [11], the Simple and Efficient TSS (SES) [12], the Four Step Search (4SS) [13] and the Diamond Search (DS) [14] are some famous examples. Such approaches have been algorithmically considered as the fastest. However, they are eventually not able to match the dynamic motion-content delivering false motion vectors (image distortions). (2) Reducing the search points, this means that the algorithm chooses, as search points, only such locations which iteratively minimize the error-function (SAD values). In this category, it is included: the Adaptive Rood Pattern Search (ARPS) [15], the Fast Block Matching Using Prediction (FBMAUPR) [16], the Block-based Gradient Descent Search (BBGD) [17] and the Neighbourhood Elimination algorithm (NE) [18]. These approaches assume that the error-function behaves monotonically, which holds well for slow-moving sequences; however, such properties do not hold true for other kind of movements in video sequences [19], yielding that the algorithms may get trapped into local minima. (3) Decreasing the computational overhead for each search point, which means the matching cost (SAD operation) is replaced by a partial or a simplify version with less complexity. The New pixel-decimation (ND) [20], the Efficient Block Matching Using Multilevel Intra and Inter-Sub-blocks [11] and the successive elimination algorithm [21]. These techniques are based on the assumption that all pixels within each block move by the same amount, while a good estimate of the motion could be obtained by using only a fraction of the pixels. However, since only a fraction of the pixels enters into the matching computation, the use of these regular sub-sampling techniques can seriously affect the accuracy of the detection of motion vectors due to the noise or illumination changes.

Alternatively, evolutionary approaches such as genetic algorithms (GA) [22] and particle swarm optimization (PSO) [23] are well known for locating potential global optimum within an arbitrary search space. In spite of such fact, only few evolutionary approaches have specifically addressed the problem of BM, such as the light-weight genetic block matching (LWG) [24], the genetic four-step search (GFSS) [25] and the PSO-BM [26]. Although these methods support an accurate identification of the motion vector, their spending times are very long in comparison to other BM techniques.

Differential Evolution (DE), introduced by Storn and Price in 1995 [27], is a novel evolutionary algorithm which is used to optimize complex continuous nonlinear functions. As a population-based algorithm, DE uses simple mutation and crossover operators to generate new candidate solutions, and applies one-to-one competition scheme to greedily decide whether the new candidate or its parent will survive in the next generation. Due to its simplicity, ease of implementation, fast convergence, and robustness, the DE algorithm has gained much attention, reporting a wide range of successful applications in the literature [28-36].

For many real-world applications, the number of calls to the objective function needs to be limited, e.g. because an evaluation is very time consuming or expensive, or because the approach requires user interaction. DE does not seem to be suited to such problems, since it usually requires many evaluations before producing a satisfying result.

The problem of excessively long fitness function calculations has already been faced in the field of evolutionary algorithms (EA), which is kwon as evolution control [37]. In an evolution control approach, the idea is to replace the costly objective function evaluation for some individuals by fitness estimates, based on an approximate model of the fitness landscape. The individuals to be evaluated, and those to be estimate, are determined based on some fixed criteria which depend on the specific properties of the used approximate model [38]. The models, used in the estimation, can be built during the actual EA run, since EA repeatedly sample the search space at different points [39]. There are certainly many possible approximation models, and several have already been used in combination with EA (e.g. polynomials [40], the kriging model [41], the feedforward neural networks, including multi-layer perceptrons [42] and radial basis-function networks [43]). These models can be either global, which make use of all available data, or local, which make use of only a small set of data around the point where the function is to be approximated. Local models, however, have a number of advantages [39]: they are well-known and established techniques, relatively fast, and take into account the intuitively most important information, the closest neighbors.

In this paper, a new algorithm based on Differential Evolution (DE) is proposed to reduce the number of search locations in the BM process. The algorithm uses a simple fitness calculation approach which is based on the Nearest Neighbor Interpolation (NNI) algorithm, in order to estimate the fitness value (SAD operation) for several candidate solutions (search locations). As a result, the approach can substantially reduce the number of function evaluations preserving the good search capabilities of DE. In comparison to other fast BM algorithms, the proposed method deploys more accurate motion vectors yet delivering competitive time rates.

The overall paper is organized as follows: Section 2 holds a brief description about the differential evolution algorithm. In Section 3, the fitness calculation strategy for solving the expensive optimization problem is presented. Section 4 provides background about the BM motion estimation issue while Section 5 exposes the final BM algorithm as a combination of DE and the NNI estimator. Section 6 demonstrates the experimental results for the proposed approach over standard test sequences as some conclusions are discussed in Section 7.

## 2. Differential evolution algorithm

The DE algorithm is a simple and direct search algorithm which is based on population and aims for optimizing global multi-modal functions. DE employs the mutation operator as to provide the exchange of information among several solutions.

There are various mutation base generators to define the algorithm type. The version of DE algorithm used in this work is known as DE/best/1/exp or ''DE1'' [27]. DE algorithms begin by initializing a population of $N_p$ and $D$-dimensional vectors considering parameter values that are randomly distributed between the pre-specified lower initial parameter bound $x_{j,\text{low}}$ and the upper initial parameter bound $x_{j,\text{high}}$ as follows:

$$x_{j,i,t} = x_{j,\text{low}} + \text{rand}(0,1) \cdot (x_{j,\text{high}} - x_{j,\text{low}});$$
$$j = 1, 2, \ldots, D; \quad i = 1, 2, \ldots, N_p; \quad t = 0. \quad (1)$$

The subscript $t$ is the generation index, while $j$ and $i$ are the parameter and particle indexes respectively. Hence, $x_{j,i,t}$ is the $j$th parameter of the $i$th particle in generation $t$. In order to generate a trial solution, DE algorithm first mutates the best solution vector $\mathbf{x}_{best,t}$ from the current population by adding the scaled difference of two vectors from the current population.

$$\mathbf{v}_{i,t} = \mathbf{x}_{best,t} + F \cdot (\mathbf{x}_{r_1,t} - \mathbf{x}_{r_2,t});$$
$$r_1, r_2 \in \{1, 2, \ldots, N_p\} \quad (2)$$

with $\mathbf{v}_{i,t}$ being the mutant vector. Indices $r_1$ and $r_2$ are randomly selected with the condition that they are different and have no relation to the particle index $i$ whatsoever (i.e., $r_1 \neq r_2 \neq i$). The mutation scale factor $F$ is a positive real number, typically less than one. Figure 1 illustrates the vector-generation process defined by Equation (2).

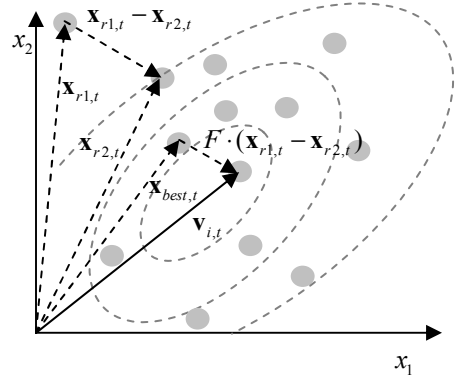

**Fig. 1.** Two-dimensional example of an objective function showing its contour lines and the process for generating **v** in scheme DE/best/1/exp from vectors of the current generation.

In order to increase the diversity of the parameter vector, the crossover operation is applied between the mutant vector $\mathbf{v}_{i,t}$ and the original individuals $\mathbf{x}_{i,t}$. The result is the trial vector $\mathbf{u}_{i,t}$ which is computed by considering element to element as follows:

$$u_{j,i,t} = \begin{cases} v_{j,i,t}, & \text{if rand}(0,1) \leq CR \text{ or } j = j_{\text{rand}}, \\ x_{j,i,t}, & \text{otherwise.} \end{cases} \tag{3}$$

with $j_{\text{rand}} \in \{1, 2, \ldots, D\}$. The crossover parameter $(0.0 \leq CR \leq 1.0)$ controls the fraction of parameters that the mutant vector is contributing to the final trial vector. In addition, the trial vector always inherits the mutant vector parameter according to the randomly chosen index $j_{\text{rand}}$, assuring that the trial vector differs by at least one parameter from the vector to which it is compared ($\mathbf{x}_{i,t}$).

Finally, a greedy selection is used to find better solutions. Thus, if the computed cost function value of the trial vector $\mathbf{u}_{i,t}$ is less or equal than the cost of the vector $\mathbf{x}_{i,t}$, then such trial vector replaces $\mathbf{x}_{i,t}$ in the next generation. Otherwise, $\mathbf{x}_{i,t}$ remains in the population for at least one more generation:

$$\mathbf{x}_{i,t+1} = \begin{cases} \mathbf{u}_{i,t}, & \text{if } f(\mathbf{u}_{i,t}) \leq f(\mathbf{x}_{i,t}), \\ \mathbf{x}_{i,t}, & \text{otherwise.} \end{cases} \tag{4}$$

Here, $f()$ represents the cost function. These processes are repeated until a termination criterion is attained or a predetermined generation number is reached.

## 3. Fitness approximation method

Evolutionary algorithms based on fitness approximation aim to find the global minimum of a given function considering only a very few number of function evaluations. In order to apply such approach, it is necessary that the objective function portrait the following conditions: [44]: (1) it must be very costly to evaluate and (2) must have few dimensions (up to five). Recently, several fitness estimators have been reported in the literature [40-43], where the function evaluation number is considerably reduced (to hundreds, dozens, or even less). However, most of these methods produce complex algorithms whose performance is conditioned to the quality of the training phase and the learning algorithm in the construction of the approximation model.

In this paper, we explore the use of a local approximation scheme, based on the nearest-neighbor-interpolation (NNI), in order to reduce the function evaluation number. The model estimates the fitness values based on previously evaluated neighboring individuals, stored during the evolution process. In each generation, some individuals of the population are evaluated with the accurate (real) objective function, while the remaining individuals' fitnesses are estimated. The individuals to be evaluated accurately are determined based on their proximity to the best fitness value or uncertainty.

*3.1 Updating individual database*

In our fitness calculation approach, during de evolution process, every evaluation or estimation of an individual produces a data point (individual position and fitness value) that is potentially taken into account for building the approximation model. Therefore, we keep all seen so far evaluations in a history array **T**, and then just select the closest neighbor to estimate the fitness value of a new individual. Thus, all data are preserved and potentially available for use, while the construction of the model is still fast since only the most relevant data points are actually used to construct the model.

*3.2 Fitness calculation strategy*

In the proposed fitness calculation scheme, most of the fitness values are estimated to reduce the calculation time in each generation. In the model, it is evaluated (using the real fitness function) those individuals that are near the individual with the best fitness value contained in **T** (rule 1). Such individuals are important, since they will have a stronger influence on the evolution process than other individuals. Moreover, it is also evaluated those individuals in regions of the search space with few previous evaluations (rule 2). The fitness values of these individuals are uncertain; since there is no close reference (close points contained in **T**) in order to calculate their estimates.

The rest of the individuals are estimated using NNI (rule 3). Thus, the fitness value of an individual is estimated assigning it the same fitness value that the nearest individual stored in **T**.

Therefore, the estimation model follows 3 different rules in order to evaluate or estimate the fitness values:

1. If the new individual (search position) $P$ is located closer than a distance $d$ with respect to the nearest individual location $L_q$ whose fitness value $F_{L_q}$ corresponds to the best fitness value stored in **T**, then the fitness value of $P$ is evaluated using the real fitness function. Figure 2a draws the rule procedure.

2. If the new individual $P$ is located longer than a distance $d$ with respect to the nearest individual location $L_q$ whose fitness value $F_{L_q}$ has been already stored in **T**, then its fitness value is evaluated using the real fitness function. Figure 2b outlines the rule procedure.

3. If the new individual $P$ is located closer than a distance $d$ with respect to the nearest individual location $L_q$ whose fitness value $F_{L_q}$ has been already stored in **T**, then its fitness value is estimated (using the NNI approach) assigning it the same fitness that $L_q$ ($F_P = F_{L_q}$). Figure 2c sketches the rule procedure.

The $d$ value controls the trade off between the evaluation and estimation of search locations. Typical values of $d$ range from 1 to 4; however, in this paper, the value of 2.5 has been selected. Thus, the proposed approach favors the exploitation and exploration in the search process. For the exploration, the estimator evaluates the true fitness function of new search locations that have been located far from the positions already calculated. Meanwhile, it also estimates those that are closer. For the exploitation, the proposed method evaluates the effective fitness function of those new searching locations that are placed near to the position with the minimum fitness value seen so far, aiming to improve its minimum.

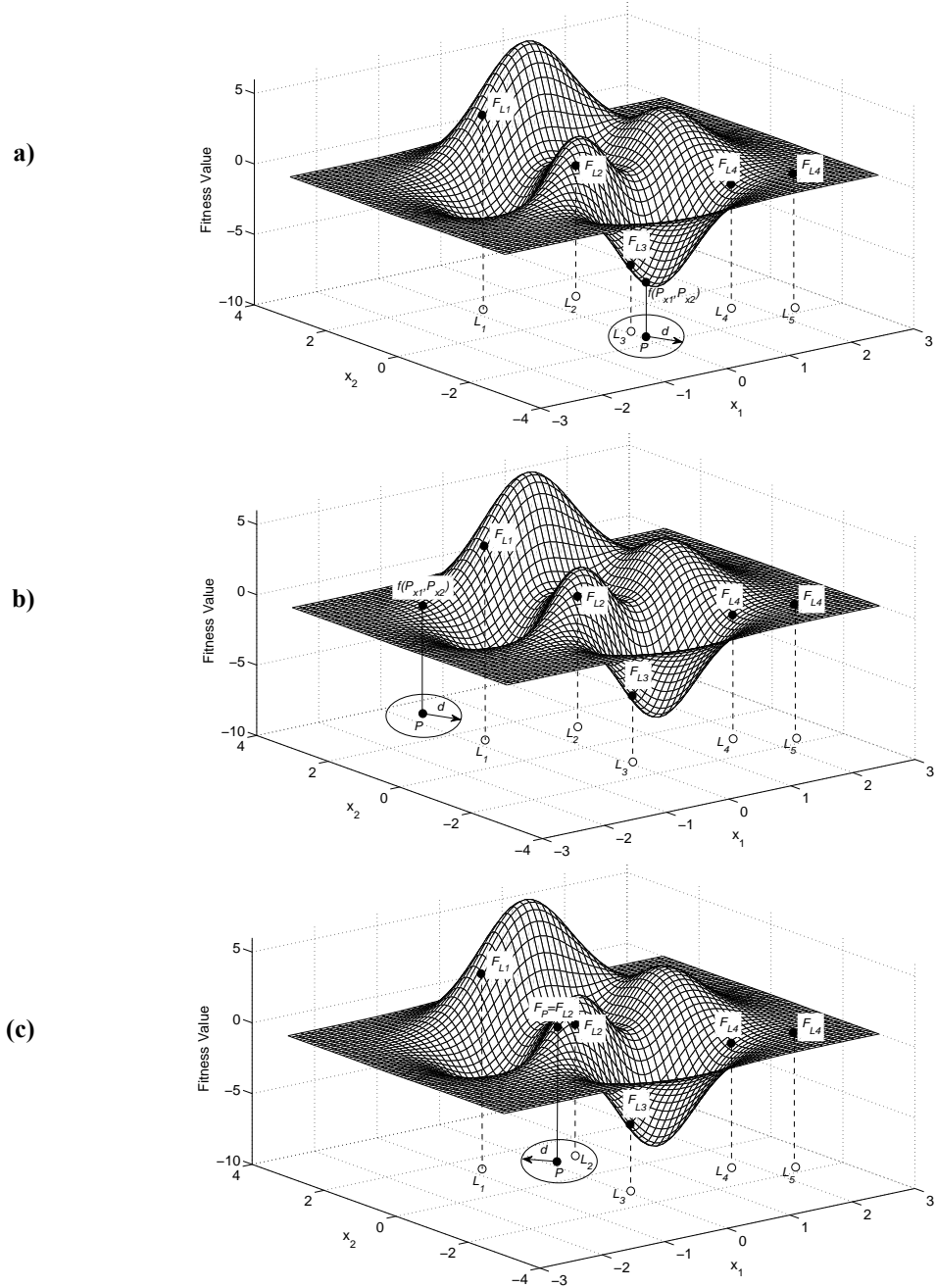

**Fig. 2.** The fitness calculation strategy. (a) According to the rule 1, the individual (search position) $P$ is evaluated, since it is located closer than a distance $d$ with respect to the nearest individual location $L_3$ whose fitness value $F_{L_3}$ corresponds to the best fitness value (minimum). (b) According to the rule 2, the search point $P$ is evaluated, as there is no close reference in its neighborhood. (c) According to the rule 3, the fitness value of $P$ is estimated by means of the NNI-estimator, assigning $F_P = F_{L_2}$

The three rules show that the fitness calculation strategy is simple and straightforward. Fig. 2 illustrates the procedure of fitness computation for a new solution (point $P$) considering the three different rules. In the problem the objective function $f$ is minimized with respect to two parameters ($x_1, x_2$). In all figures (Figs. 2(a), (b) and (c)) the individual database array **T** contains five different elements ($L_1, L_2, L_3, L_4, L_5$) with their corresponding fitness values ($F_{L_1}, F_{L_2}, F_{L_3}, F_{L_4}, F_{L_5}$). Figures 2(a) and (b) show the fitness evaluation ($f(x_1, x_2)$) of the new solution $P$ following the rule 1 and 2 respectively, whereas Fig. 2(c) present the fitness estimation of $P$ using the NNI approach considered by rule 3.

*3.3 Proposed optimization DE method*

In this section, it has been proposed a fitness calculation approach in order to accelerate the DE algorithm. Only the fitness calculation scheme shows difference between the conventional DE and the enhanced one. In the modified DE, only some individuals are actually evaluated (rules 1 and 2) in each generation. The fitness values of the rest are estimated using the NNI-approach (rule 3). The estimation is executed using the individual database (array **T**).

Fig. 3 shows the difference between the conventional DE and the modified one. In the Figure, it is clear that two new blocks have been added, the fitness estimation and the updating individual database. Both elements, together with the actual evolution block, represent the fitness calculation strategy presented in this section. As a result, the DE approach can substantially reduce the number of function evaluations preserving the good search capabilities of DE.

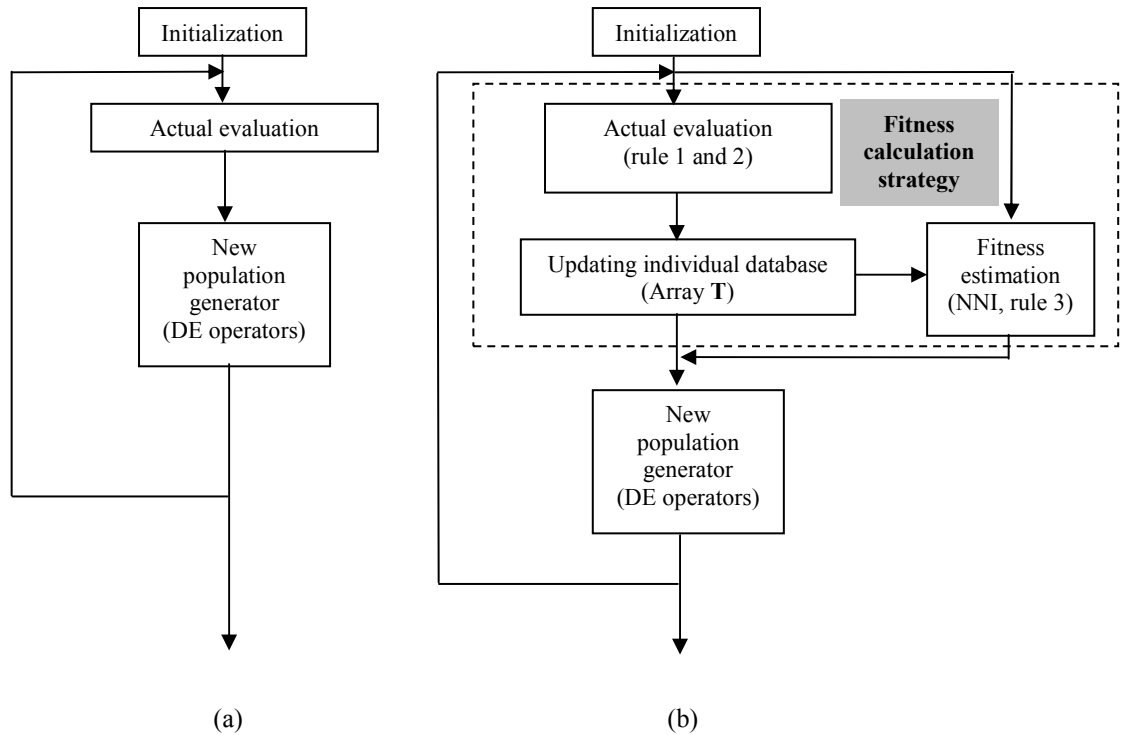

**Fig. 3.** Differences between the conventional DE and the modified DE. (a) Conventional DE and (b) DE algorithm included the fitness calculation strategy

## 4. Motion estimation and Block matching

For motion estimation, in a BM algorithm, the current frame of an image sequence $I_t$ is divided into non-overlapping blocks of *N*x*N* pixels. For each template block in the current frame, the best matched block within a search window of size (2*W*+1)x(2*W*+1) in the previous frame $I_{t-1}$ is determined, where *W* is the maximum allowed displacement. The position difference between a template block in the current frame and the best matched block in the previous frame is called the Motion Vector (MV) (see Fig. 4).

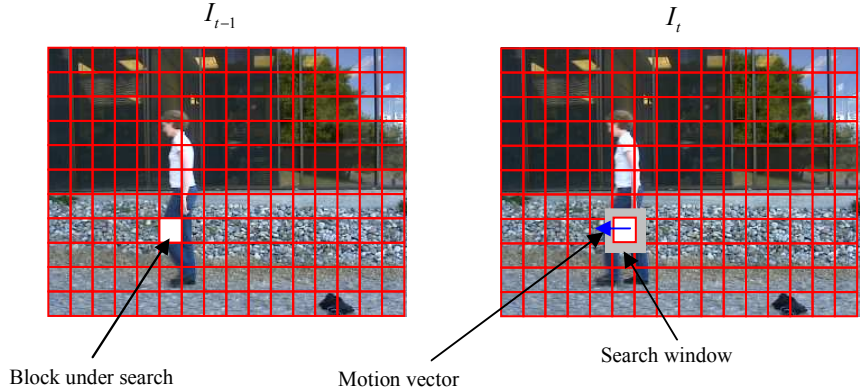

Block under search     Motion vector     Search window

**Fig. 4.** Block Matching procedure.

The most commonly used criterion for BM algorithms is the sum of absolute difference (SAD), which is defined in Eq.(5), between a template block at position (*x*, *y*) in the current frame and the candidate block at position $(x+\hat{u}, y+\hat{v})$ in the previous frame $I_{t-1}$.

$$\text{SAD}(\hat{u},\hat{v}) = \sum_{j=0}^{N-1}\sum_{i=0}^{N-1} \left| g_t(x+i, y+j) - g_{t-1}(x+\hat{u}+i, y+\hat{v}+j) \right| \tag{5}$$

where $g_t(\cdot)$ is the gray value of a pixel in the current frame $I_t$ and $g_{t-1}(\cdot)$ is the gray level of a pixel in the previous frame $I_{t-1}$. Therefore, the MV in $(u,v)$ is defined as follows:

$$(u,v) = \arg\min_{(u,v)\in S} \text{SAD}(\hat{u},\hat{v}) \tag{6}$$

where $S = \{(\hat{u},\hat{v}) \mid -W \leq \hat{u},\hat{v} \leq W \text{ and } (x+\hat{u}, y+\hat{v}) \text{ is a valid pixel position } I_{t-1}\}$.

The FSA is the most robust and accurate method to find the MV. It tests all possible candidate blocks from $I_{t-1}$ within the search area to find the block with minimum SAD. For the maximum displacement of *W*, the FSA requires $(2W+1)^2$ search points. For instance, if the maximum displacement *W* is 7, the total search points are 225. Each SAD calculation requires $2N^2$ additions as the total number of addition for the FSA to match a 16×16 block is 130560. The computational requirement makes difficult the application of FSA in real time applications.

## 5. BM algorithm based on DE with the estimation strategy

FSA finds the global minimum (the accurate MV), considering all locations within the search space *S*. Nevertheless, this approach has a high computational cost for practical use. To overcome such a problem, many fast algorithms have been developed despite their precision is poorer than the FSA. A better BM algorithm should spend less computation time on searching and obtaining accurate motion vectors (MVs).

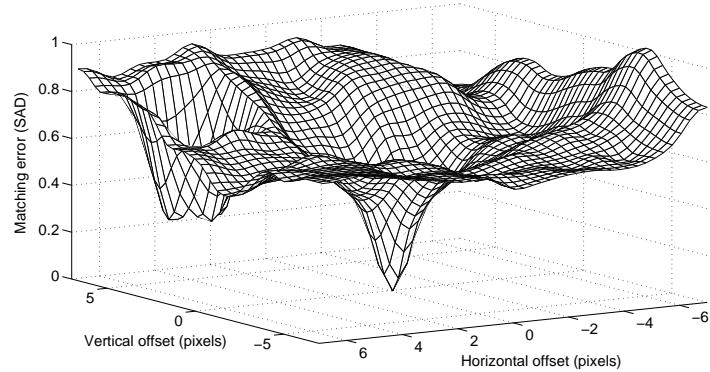

**Fig. 5.** Common non-uni-modal error surface with multiple local minimum error points.

The BM algorithm, proposed in this paper, has the velocity of the fastest algorithms and a precision similar to the FSA approach. Because most of the fast algorithms use a regular search pattern or assume a characteristic error function (uni-modal) for searching the motion vector, they may get trapped into local minima, considering that, for many cases (complex motion sequences), the uni-modal error assumption is not longer valid. Fig. 5 shows a typical error surface (SAD values) which has been computed around the search window considering a fast-moving sequence. On the other hand, the proposed BM algorithm uses a non-uniform search pattern for locating global minimum distortion. Under the effect of the DE operators, the search locations vary from generation to generation, avoiding to get trapped into a local minimum. Besides, since the proposed algorithm uses a fitness calculation strategy in order to reduce the evaluation of the SAD values, it uses few search positions.

In the algorithm, the search space $S$ consists of a set of 2-D motion vectors $\hat{u}$ and $\hat{v}$ representing the *x* and *y* components of the motion vector, respectively. The particle is defined as:

$$P_i = \{\hat{u}_i, \hat{v}_i \mid -W \leq \hat{u}_i, \hat{v}_i \leq W\} \tag{7}$$

where each particle *i* represents a possible motion vector. In this paper, the maximum offset is *W*=7 pixels.

*5.1 Initial population*

The first step in DE optimization is to generate an initial group of particles. The standard literature of evolutionary algorithms generally suggests the use of random solutions as the initial population, assuming the absence of knowledge about the problem [45]. On the other hand, Li [46] and Xiao [47] demonstrated that the use of solutions generated through some domain knowledge (i.e., non-random solutions) to set the initial population can significantly improve its performance. In order to obtain appropriate initial solutions (based on knowledge), an analysis over the motion vector distribution should be conducted. After considering several sequences (see Table 1 and Fig. 9), it can be seen that 98% of the MVs are found to lie at the origin of the search window for a slow-moving sequence such as the one at *Container*, whereas complex motion sequences, such as the *Carphone* and the *Foreman* examples, have only 53.5% and 46.7% of their MVs in the central search region. The *Stefan* sequence, showing the most complex motion content, has only 36.9%. Figure 6 shows the surface of the MV distribution for the F*oreman* and the *Stefan*. On the other hand, although it is less evident, the MV distribution of several sequences shows small peaks at some locations lying away from the center as they are contained inside a rectangle that is shown in Fig. 6(b) and 6(d) by a white overlay. Real-world moving sequences concentrate most of the MVs under a limit due to the motion continuity principle [41]. Therefore, in this paper, initial solutions are selected from five fixed locations which represent locations showing the higher concentration in the MV distribution, just as it is shown by Figure 7.

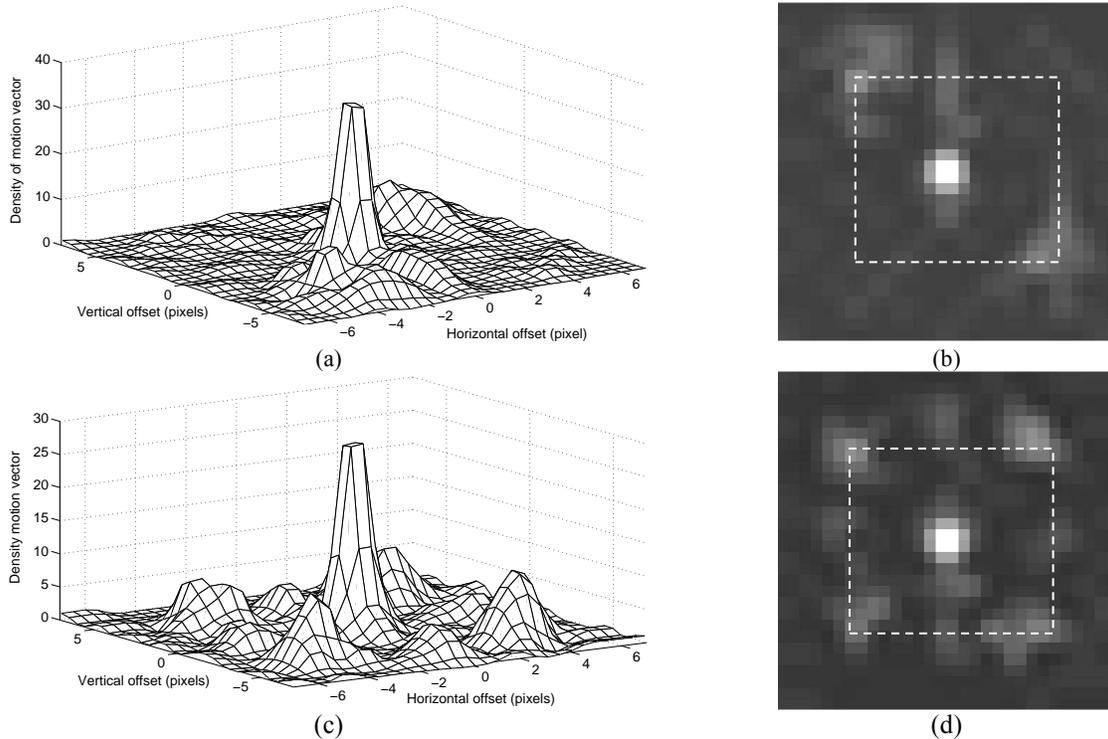

**Fig. 6.** Motion vector distribution for *Foreman* and Stefan sequences. (a)-(b) MV distribution for the *Foreman* sequence. (c)-(d) MV distribution for the *Stefan* sequence.

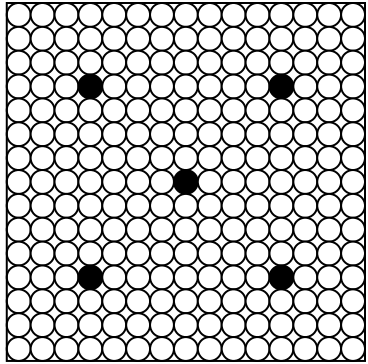

**Fig. 7.** Fixed pattern of five elements used as initial solutions.

*5.2 The DE-BM algorithm*

The goal of our BM-approach is to reduce the number of evaluations of the SAD values (real fitness function) without loosing performance on achieving an acceptable solution. The DE-BM method is presented below:

**Step 1:**   Set the DE parameters (*F*=0.25, *CR*=0.8, see Section 2).

**Step 2:**   Initialize the population of 5 individuals using the pattern shown in Fig. 7 and the individual database array **T**, without elements.

| | |
|---|---|
| **Step 3:** | Compute the fitness values of each individual using the fitness calculation strategy presented in Section 3. Since all individuals of the initial population fulfil the conditions of rule 2, they are evaluated with the real fitness function (calculating the real SAD values). |
| **Step 4:** | Update the new evaluations in the individual database array **T**. |
| **Step 5:** | Generate a new population of five individuals (trial population) considering the DE operators of mutation Eq. (2) and crossover Eq. (3). |
| **Step 6:** | Compute the fitness values of each individual using the fitness calculation strategy presented in Section 3. |
| **Step 7:** | Update the new evaluations (rule 1 and 2) or estimations (rule 3) in the individual database array **T**. |
| **Step 8:** | Select the fittest element between each individual and its corresponding trial counterpart according to Eq. (4) in order to obtain the final individual for the next generation. |
| **Step 9:** | If seven iterations have not been reached, then go back to Step 5; otherwise the best individual ($\hat{u}_{best}$, $\hat{v}_{best}$) of the final population is considered the MV. |

The proposed DE-BM algorithm uses 40 different individuals (search locations) during the complete optimization process. However, only from 7 to 18 search locations are evaluated using the real fitness function (SAD evaluation) while the remaining positions are just estimated. Figure 8 shows two search-patterns examples that have been generated by the DE-BM approach. Such patterns exhibit the evaluated search-locations (rule 1 and 2) in white-cells whereas the minimum location is marked in black. The grey-cells represent the cells that were estimated (rule 3) or not visited during the optimization process.

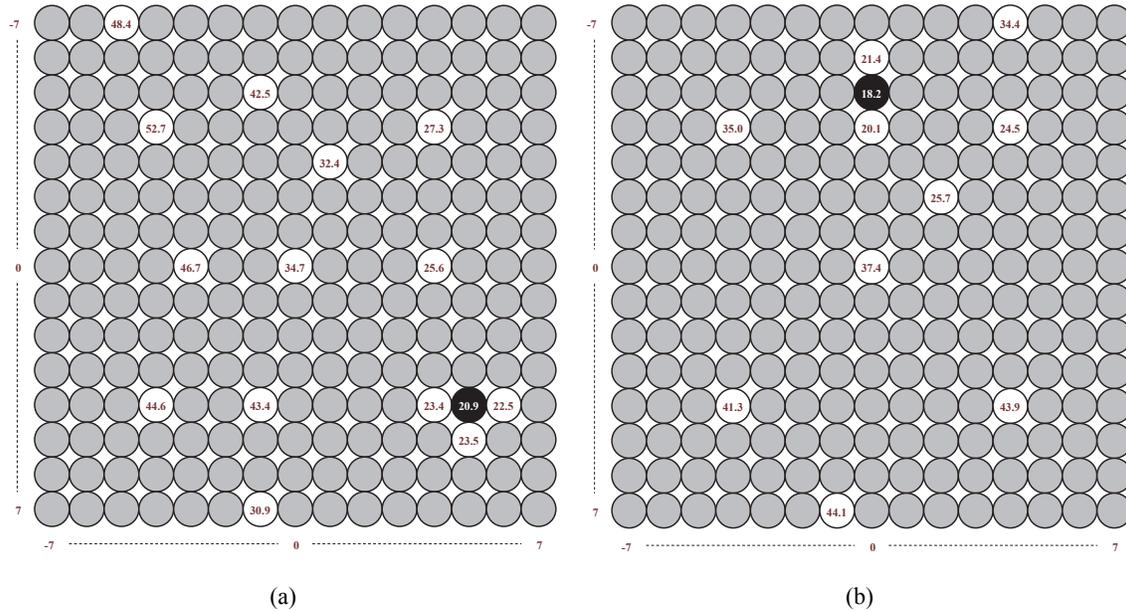

(a)        (b)

**Fig. 8.** Search-patterns generated by the DE-BM algorithm. (a) Pattern with solution $\hat{u}^1_{best} = 5$ and $\hat{v}^1_{best} = 4$ .(b) Pattern with solution $\hat{u}^2_{best} = -5$ and $\hat{v}^2_{best} = 0$ .

## 6. Experimental results

This section presents the results of comparing the proposed DE-BM algorithm to other existing block matching algorithms. The simulations have been performed over the luminance component of popular video sequences that are listed in Table 1. Such sequences consist of different degrees and types of motion including

QCIF (176x144), CIF (352x288) and SIF (352x240) respectively. The first four sequences are *Container*, *Carphone*, *Foreman* and *Akiyo* in QCIF format. The next two sequences are *Stefan* in CIF format and *Tennis* in SIF format. Among these sequences, Container has gentle, smooth and low motion change and consists mainly of stationary and quasi-stationary blocks. *Carphone*, *Foreman* and *Akiyo* have moderately complex motion getting a ''medium'' category regarding its motion content. Rigorous motion which is based on camera panning with translation and complex motion content can be found in the sequences of *Stefan* and *Tennis*. Figure 9 shows a sample frame from each sequence.

Each picture frame is partitioned into macro-blocks with the sizes of 16x16 ($N$=16) pixels for motion estimation where the maximum displacement within the search range is ±7 pixels in both the horizontal and the vertical directions.

In order to compare the performance of the DE-BM approach, different search algorithms such as FSA, TSS [10], 4SS [13], NTSS [11], BBGD [17], DS [14], NE [18], ND [20], LWG [24], GFSS [25] and PSO-BM [26] have been all implemented in our simulations. For comparison purposes, all six video sequences in Fig. 8 have been used. All simulations are performed on a Pentium IV 3.2 GHz PC with 1GB of memory.

In the comparison, three relevant performance indexes have been considered: the coding quality and the search efficiency.

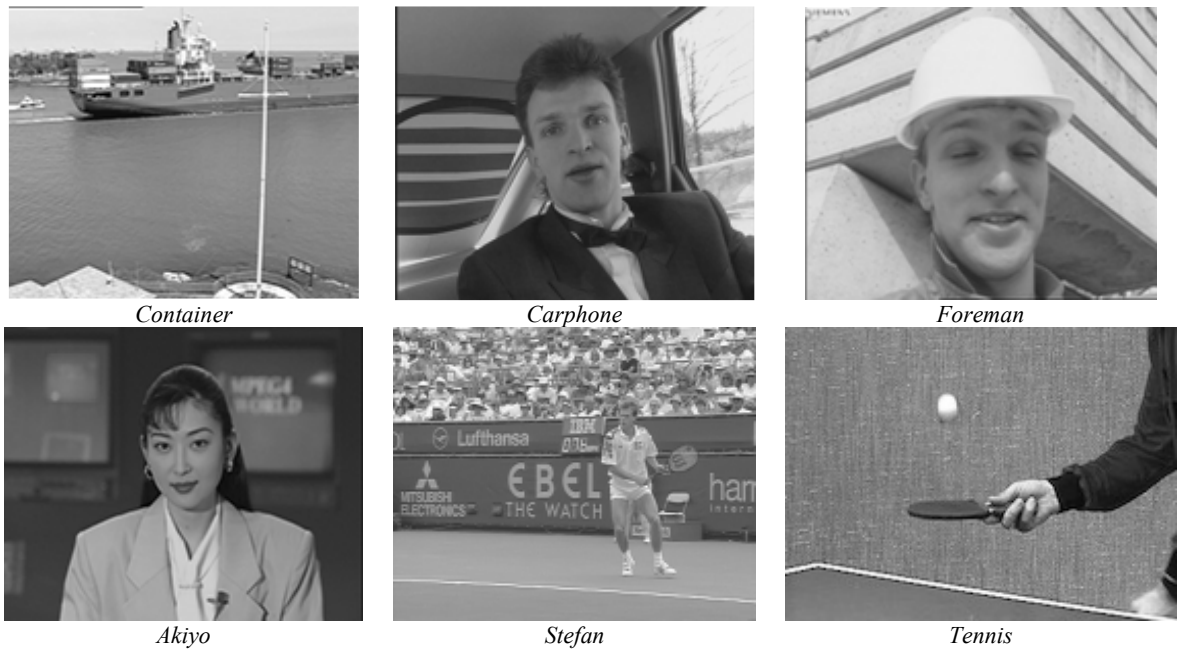

| *Container* | *Carphone* | *Foreman* |
| *Akiyo* | *Stefan* | *Tennis* |

**Fig. 9.** Test video sequences.

Table 1. Test sequences used in the comparison test.

| Sequence | Format | Total frames | Motion type |
|---|---|---|---|
| *Container* | QCIF(176x144) | 299 | Low |
| *Carphone* | QCIF(176x144) | 381 | Medium |
| *Foreman* | QCIF(352x288) | 398 | Medium |
| *Akiyo* | QCIF(352x288) | 211 | Medium |
| *Stefan* | CIF(352x288) | 89 | High |
| *Tennis* | SIF(352x240) | 150 | High |

*6.1 Coding quality*

First, all algorithms are compared in terms of their coding quality. The coding quality is characterized by the Peak-Signal-to-Noise-Ratio (PSNR) value which indicates the reconstruction quality when the motion vectors, computed by a BM approach, are used. In PSNR, the signal is the original data frames whereas the noise is the error introduced by the calculated motion vectors. The PSNR is defined as

$$\text{PSNR} = 10 \cdot \log_{10}\left(\frac{255^2}{MSE}\right) \quad (8)$$

where *MSE* is the mean square between the original frames and those compensated by the motion vectors. Additionally, as an alternative performance index, it is used in the comparison the PSNR degradation ratio ($D_{PSNR}$). This ratio expresses in percentage (%) the level of mismatch between the PSNR of a BM approach and the PSNR of the FSA which is considered as reference. $D_{PSNR}$ is defined as

$$D_{PSNR} = -\left(\frac{\text{PSNR}_{FSA} - \text{PSNR}_{BM}}{\text{PSNR}_{FSA}}\right) \cdot 100\% \quad (9)$$

**Table 2.** PSNR values and $D_{PSNR}$ comparison for all BM methods

| Algorithm | Container | | Carphone | | Foreman | | Akiyo | | Stefan | | Tennis | | Total Average ($D_{PSNR}$) |
|---|---|---|---|---|---|---|---|---|---|---|---|---|---|
| | PSNR | $D_{PSNR}$ | PSNR | $D_{PSNR}$ | PSNR | $D_{PSNR}$ | PSNR | $D_{PSNR}$ | PSNR | $D_{PSNR}$ | PSNR | $D_{PSNR}$ | |
| FSA | 43.18 | 0 | 31.51 | 0 | 31.69 | 0 | 29.07 | 0 | 25.95 | 0 | 35.74 | 0 | 0 |
| TSS | 43.10 | -0.20 | 30.27 | -3.92 | 29.37 | -7.32 | 26.21 | -9.84 | 21.14 | -18.52 | 30.58 | -14.42 | -9.03 |
| 4SS | 43.12 | -0.15 | 30.24 | -4.01 | 29.34 | -7.44 | 26.21 | -9.84 | 21.41 | -17.48 | 30.62 | -14.32 | -8.87 |
| NTSS | 43.12 | -0.15 | 30.35 | -3.67 | 30.56 | -3.57 | 27.12 | -6.71 | 22.52 | -13.20 | 31.21 | -12.65 | -6.65 |
| BBGD | 43.14 | -0.11 | 31.30 | -0.67 | 31.00 | -2.19 | 28.10 | -3.33 | 25.17 | -3.01 | 33.17 | -7.17 | -2.74 |
| DS | 43.13 | -0.13 | 31.26 | -0.79 | 31.19 | -1.59 | 28.00 | -3.70 | 24.98 | -3.73 | 33.98 | -4.92 | -2.47 |
| NE | 43.15 | -0.08 | 31.36 | -0.47 | 31.23 | -1.47 | 28.23 | -2.89 | 25.22 | -2.81 | 33.88 | -5.19 | -2.15 |
| ND | 43.15 | -0.08 | 31.35 | -0.50 | 31.20 | -1.54 | 28.21 | -2.96 | 25.21 | -2.86 | 33.79 | -5.43 | -2.22 |
| LWG | 43.16 | -0.06 | 31.40 | -0.36 | 31.31 | -1.21 | 28.55 | -1.80 | 25.41 | -2.09 | 33.97 | -4.95 | -1.74 |
| GFSS | 43.15 | -0.06 | 31.38 | -0.40 | 31.29 | -1.26 | 28.32 | -2.58 | 25.34 | -2.36 | 33.87 | -5.23 | -1.98 |
| PSO-BM | 43.15 | -0.07 | 31.39 | -0.38 | 31.27 | -1.34 | 28.33 | 2.55 | 25.39 | -2.15 | 33.91 | -5.11 | -1.93 |
| DE-BM | 43.17 | -0.04 | 31.47 | -0.13 | 31.51 | -0.58 | 28.98 | -1.00 | 25.85 | -0.78 | 34.77 | -4.27 | -1.13 |

Table 2 shows the comparison of the PSNR values and the PSNR degradation ratios ($D_{PSNR}$) among the different algorithms, considering the six image sequences presented in Fig. 8. As it can be seen in the case of the slow-moving sequence *Container*, the PSNR values (the $D_{PSNR}$ ratios) of all BM algorithms are similar. For the medium motion content sequences such as *Carphone*, *Foreman* and *Akiyo*, the approaches consistent of fixed patterns (TSS, 4SS and NTSS) exhibit the worst PSNR values (high $D_{PSNR}$ ratios) except for the DS algorithm. On the other hand, the BM methods that use evolutionary algorithms (LWG, GFSS, PSO-BM and DE-BM) present the lowest $D_{PSNR}$ ratios, only one step under the FSA approach which is considered as reference. Finally, the approaches based on error-function minimization (BBGD and NE) and pixel-decimation (ND) posses a medium performance. For the high motion sequences such as *Stefan* and *Tennis*, the same conclusions, as the medium motion content sequences, can be observed. Since the motion content of these sequences in complex (producing error surfaces with more than one minimum), the performance, in general, becomes worst for most of the algorithms. However, the PSNR values (or $D_{PSNR}$ ratios) of the DS and the DE-BM approaches maintain a better performance. As a summary of the coding quality performance, the last column of Table 2 presents the average PSNR degradation ratio ($D_{PSNR}$) obtained over all sequences. According to these values, the proposed DE-BM method is superior to any other approach (due to the computation complexity, the FSA is considered just as a reference).

Fig. 9(a)-(b) shows the comparison of the frame-wise coding performance for the *Akiyo* and *Tennis* sequences. Since the algorithms FSA, DE-BM, LWG, BBGD, NE and DS obtained good PSNR values in the coding performance analysis, they are only considered in the comparison. According to these graphs, the coding quality of the DE-BM is better than the other algorithms (FSA, LWG, BBGD, NE and DS) whose PSNR values fluctuates heavily for some regions in the video sequences. The PSNR values of the DE-BM algorithm are only lightly under the FSA approach which is used as reference. For a high motion sequence such as *Tennis*, we present a comparative result for frame-wise PSNR in Fig. 9b. Results exhibit a similar pattern as the one discussed for the *Akiyo* sequence.

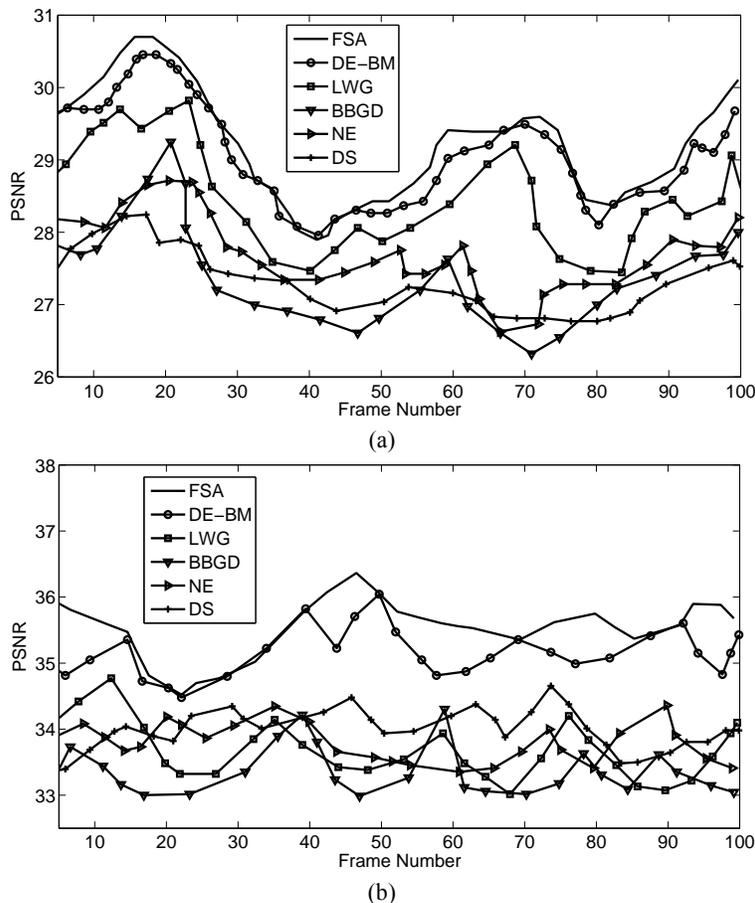

**Fig. 9.** Frame-wise performance comparison between different BMAs on sequence (a) Akiyo and (b) Tennis, considering 100 frames.

*6.2 Search efficiency*

The search efficiency is used, in this paper, as a measurement of computational complexity. The search efficiency is calculated by counting the average number of search points (or the average number of SAD computations) for a MV estimation. In Table 3, the search efficiency is compared. Only a step above FSA, evolutionary algorithms LWG, GFSS and PSO-BM hold the highest number of search points per block. On the contrary, the proposed DE-BM algorithm maintains a similar performance to BBGDS and DS representing the fastest approaches. From the data shown in Table 3, the average number of search locations, corresponding to the DE-BM method, represents the number of SAD evaluations (the number of SAD estimations are not considered). Additionally in the last two columns of Table 3, it is presented the number of search locations averaged over the six considered sequences and the rank occupied for each approach. According to these values, the proposed DE-BM method is ranking in the second place, a step under BBGDS. The average number of search points checked by the DE-BM algorithm is only from 9.2 to 16.8, which are 4% and 7.4% that of the FSA method. These results demonstrate that our approach can significantly reduce the number of search points. Hence, the DE-BM algorithm is at least equal to the other fast methods in terms of the reduction of the number of search points.

**Table 3.** Average number of search points per block for all ten BM methods.

| Algorithm | Container | Carphone | Foreman | Akiyo | Stefan | Tennis | Total Average | Rank |
|---|---|---|---|---|---|---|---|---|
| FSA | 225 | 225 | 225 | 225 | 225 | 225 | 225 | 12 |
| TSS | 25 | 25 | 25 | 25 | 25 | 25 | 25 | 7 |
| 4SS | 19 | 25.5 | 24.8 | 27.3 | 29.3 | 31.5 | 26.3 | 8 |
| NTSS | 17.2 | 21.8 | 22.1 | 23.5 | 25.4 | 26.1 | 22.6 | 6 |
| BBGD | 8.1 | 11.5 | 12.5 | 10.2 | 15.2 | 17.1 | 12.43 | 1 |
| DS | 7.5 | 12.5 | 13.4 | 11.8 | 16.2 | 17.5 | 13.15 | 3 |
| NE | 11.7 | 13.8 | 14.2 | 14.5 | 19.2 | 20.2 | 15.6 | 5 |
| ND | 10.8 | 13.4 | 13.8 | 14.1 | 18.4 | 19.1 | 14.9 | 4 |
| LWG | 75 | 75 | 75 | 75 | 75 | 75 | 75 | 11 |
| GFSS | 60 | 60 | 60 | 60 | 60 | 60 | 60 | 10 |
| PSO-BM | 32.5 | 48.5 | 48.1 | 48.5 | 52.2 | 52.2 | 47 | 9 |
| DE-BM | 9.2 | 12.2 | 12.2 | 12.5 | 16.1 | 16.8 | 13.14 | 2 |

**7. Conclusions**

In this paper, a new algorithm based on Differential Evolution (DE) is proposed to reduce the number of search locations in the BM process. In order to save computational time, the approach combines the traditional DE with a fitness estimation strategy that decides which search locations (individuals) can be estimated or actually evaluated. As a result, the approach can substantially reduce the number of function evaluations, yet preserving the good search capabilities of DE.

The used fitness calculation strategy estimates the SAD (fitness) value of search locations using previously evaluated neighboring locations which have been visited during the evolution process. In the strategy, those positions close to the location with the best fitness value (seen so-far), are evaluated by using the actual fitness function. Similarly, it is also evaluated those positions lying in regions of the search space with no previous evaluations. The remaining search positions are estimated assigning them the same fitness value that the nearest known location. By the use of such fitness estimation method, the SAD value of only very few search positions are actually evaluated whereas the rest is just estimated.

Since the proposed algorithm does not consider any fixed search pattern or any other movement assumption, a high probability for finding the true minimum (accurate motion vector) is expected regardless of the movement complexity contained in the video sequence. Therefore, the chance of being trapped into a local minimum is reduced in comparison to other BM algorithms.

The performance of DE-BM has been compared to other existing BM algorithms (FSA, TSS [10], 4SS [13], NTSS [11], BBGD [17], DS [14], NE [18], ND [20], LWG [24], GFSS [25] and PSO-BM [26]) considering different sequences which present a great variety of formats and movement types. Experimental results demonstrate the high performance of the proposed method in terms of computational complexity and coding efficiency, reducing the number search points about 95% and preserving a negligible degradation ratio $D_{PSNR}$ of -1.13, respectively.